# AN EXPERIMENTAL METHOD TO DIRECTLY OBTAIN A PHOTOGENERATED CURRENT AT EACH JUNCTION WITHIN THE TRIPLE JUNCTION SOLAR CELL

Nelson Veissid and Ricardo Augusto Santos de Abreu

Laboratório Associado de Sensores e Materiais / CTE / INPE

CP 515, CEP 12245-970, São José dos Campos, São Paulo-Brazil

FAX: (++5512) 3945-6717, e-mail: veissid@las.inpe.br

**Abstract** – In this paper we will report an experimental approach for directly determining the photogenerated current at individual junctions within the triple junction solar cell. The method is based on the measurement of different current-voltage curves; one with the usual illumination system, normally AM0 or AM1.5G, and the other curves with a light source, having a different spectral emission. The fitting of all current-voltage curves simultaneously, not individually, permit us to determine the photogenerated current of three junctions. It is necessary to have three characteristics or more. An example, with experimental current-voltage data points, is shown in this work.

The main advantage of this new technique is its simplicity and its easy implementation at any laboratory with an electrical characterization solar cell system and a solar simulator. The method has been tested over a large quantity of TJ solar cells.

## 1 - INTRODUCTION

Satellites are powered by solar cells and, recently, triple junction solar cells have received considerable attention because of their high performance and radiation resistance (Cotal et al., 2000). Unfortunately, the short circuit current of a multi junction solar cell is limited by the junction that has the smaller photogenerated current. This phenomenon occurs because the junctions are connected in series, and only the smaller short circuit current is visible on measurement of current-voltage characteristic. To obtain the photogenerated current of other junctions, the electric characterization becomes very complex; and several methods, including spectral response analysis, are used to study the behavior of each junction as a function of temperature, light intensity or radiation damage (Sumita et al., 2003).

The best algorithm, for the determination of the set values parameters of the current-voltage (IxV) characteristic equation, is using the full IxV curve (Veissid et al., 1995). The electric circuit equivalent of a homojunction solar cell defines the equation below for the current versus voltage characteristic.

$$I = I_L - I_{01}\left\{\exp\left[\frac{q(V + I.R_S)}{kT}\right] - 1\right\} - I_{02}\left\{\exp\left[\frac{q(V + I.R_S)}{AkT}\right] - 1\right\} - \frac{V + I.R_S}{R_P} \quad (1)$$

In this work, an almost ideal solar cell was considered, therefore the second exponential and shunt resistance were omited. Eq. (2) shows the simplified curve of I=f(V) and Eq. (3) shows the voltage as a function of current, taking into consideration the condition when the difference between photogenerated current and the measured current is higher than the saturation current.

$$I = I_L - I_0\left\{\exp\left[\frac{q(V + I.R_S)}{kT}\right] - 1\right\} \quad (2)$$

$$V = \frac{k.T}{q}\ln\left(\frac{I_L - I}{I_0}\right) - I.R_S \quad (3)$$

The IxV characteristic of the triple junction solar cell can be understood as a device with three homojunctions connected in series. Hamakawa and Okamoto (1984) have defined the IxV curve to the multijunction solar cell and their equation can be adapted to TJ solar cell and is given by Eq. (4).

$$V = \frac{k.T}{q}\{\ln[\frac{(I_{L1} - I).(I_{L2} - I).(I_{L3} - I)}{I_{01}.I_{02}.I_{03}}]\} - I.R_S \tag{4}$$

**2 – EXPERIMENTAL METHOD**

The experimental apparatus consists of a MHG (multi halogen gas, 545W) lamp with two sets of red PAR lamps and, additional, mirrors-lens to achieve power and homogeneity at an area of 10cmx10cm. Three illuminated IxV curves are measured using an automatic data acquisition system. The first measurement is obtained using only the MHG lamp, the second is with the MHG lamp with an additional set of red PAR lamp end; the last IxV curve is measured with all lamps turned on. Fig. 1 shows the curves of the AM0 spectral and the spectral irradiance of the lamps used in this work, which were taken with a spectrometer.

Eq. 5, obtained from Eq. 4, establishes the behavior of three curves simultaneously.

$$V = \frac{k.T}{q}\{\ln[\frac{(I_{L1} + z.\Delta I_{L1} - I).(I_{L2} + z.\Delta I_{L2} - I).(I_{L3} + z.\Delta I_{L3} - I)}{I_{01}.I_{02}.I_{03}}]\} - I.R_S \tag{5}$$

The parameter z refers to increment in the photogenerated current due to the red par lamp and it has the following values: z=0 with only MHG lamp (no red par lamp), z=1 with one set of red par lamp more MHG lamp and z=2 to all lamps working.

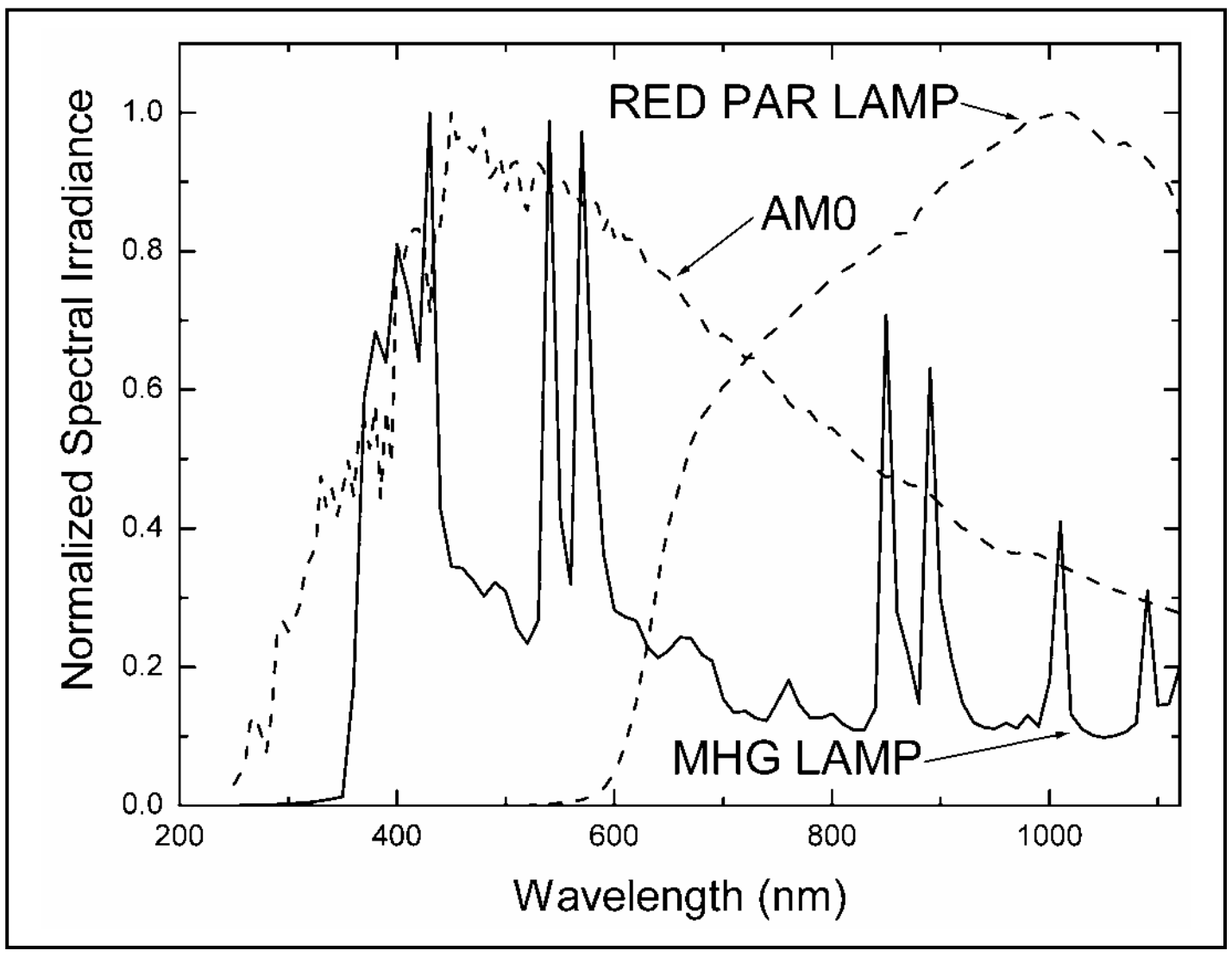

Fig. 1 – AM0 and the lamps of this work spectral irradiances curves.

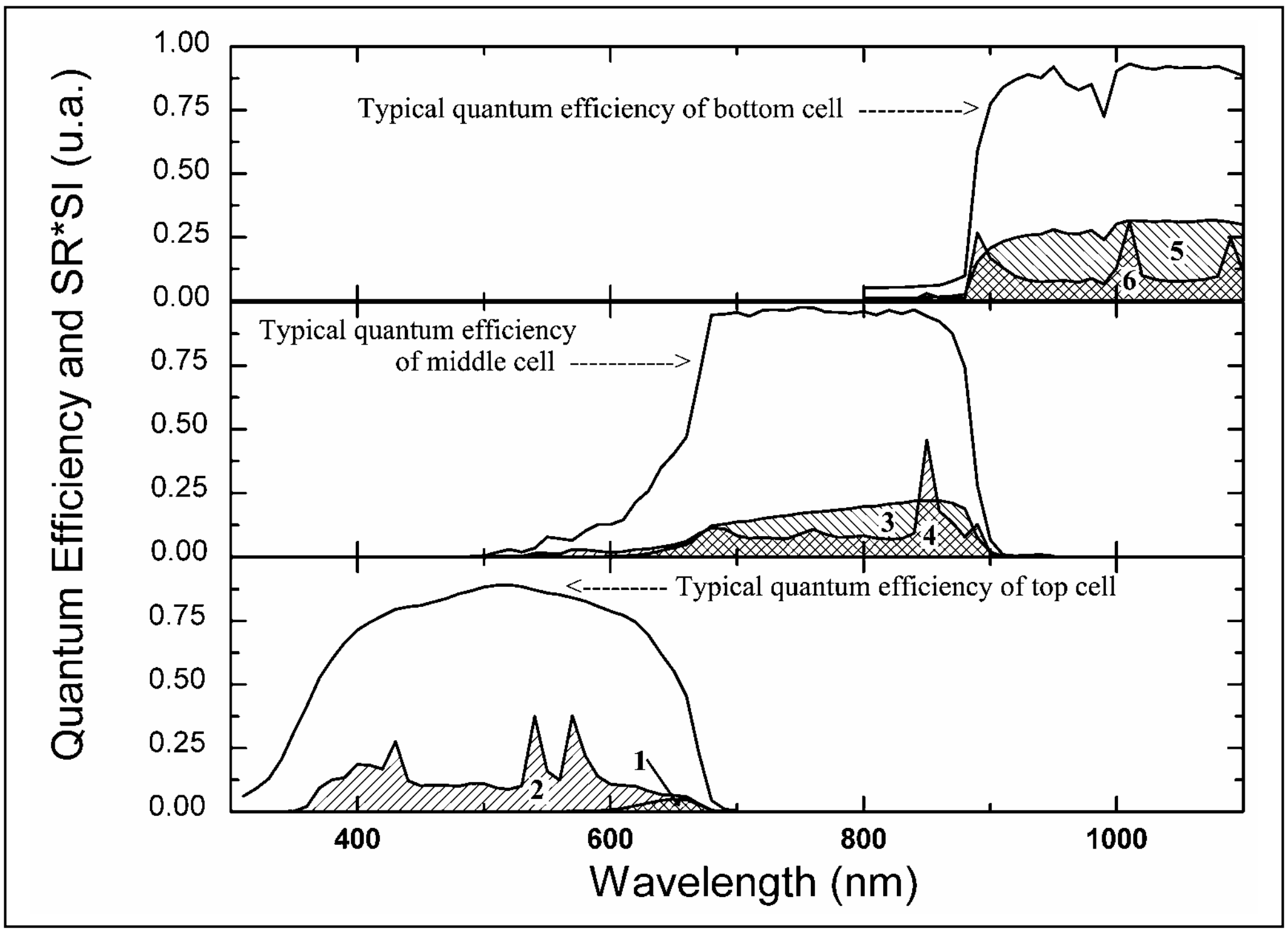

Fig. 2 – Typical quantum efficiencies (QE) of the junctions in the TJ solar cell. Curves of the product the typical spectral response with the spectral irradiance shown in Fig. 1. The areas under the curves indicate an estimate of photogenerated current.

Fig. 2 permits us to analyze qualitatively the photogenerated current in the junctions. The following intuitive considerations are possible to extract from this figure:

A-In the top cell, the increment produced with red par lamp (area 1) is around one order of magnitude smaller than the photogenerated current of the MHG lamp (area 2).

B-In the middle cell, the increment (area 3) is approximately equal to the area 4.

C-In the bottom cell, the increment (area 5) is several times higher than the photogenerated current of the MHG lamp (area 6).

**5 – RESULTS**

Karam et al. (1999, 2001) have extensively described the fabrication and characterization of the TJ solar cell. The device used in this work has the following characteristics (according to the manufacturer): Area=27cm$^2$, $V_{OC}$=2.542V, $I_{SC}$=0.447A, $V_{MAX}$=2,278V, $I_{MAX}$=0.435A and efficiency=26.4%. Table 1 shows the IxV values measured in this TJ solar cell at a temperature of 25°C, where the first column are the values with only the MHG lamp (z=0), the second is with MHG lamp with an additional  of red par lamp (z=1) and the third is with all lamps (z=2). The experimental voltage accuracy is 1 mV. The voltage range is higher than 2.3V to avoid the influence of the shunt resistance and other effects that were not considered in the mathematical model used in this work.

Table 1 – Experimental and fitted data points of the current-voltage characteristics. The parameter z defines the illumination condition.

| z=0 | | | z=1 | | | z=2 | | |
|---|---|---|---|---|---|---|---|---|
| $V_{EXP}$(V) | $I_{EXP}$(A) | $V_{fitted}$(V) | $V_{EXP}$(V) | $I_{EXP}$(A) | $V_{fitted}$(V) | $V_{EXP}$(V) | $I_{EXP}$(A) | $V_{fitted}$(V) |
| 2.370 | 0.12069 | 2.3699 | 2.408 | 0.19409 | 2.4067 | 2.429 | 0.23169 | 2.4284 |
| 2.393 | 0.10434 | 2.3933 | 2.428 | 0.17217 | 2.4285 | 2.448 | 0.20923 | 2.4495 |
| 2.411 | 0.08833 | 2.4108 | 2.444 | 0.15207 | 2.4446 | 2.463 | 0.18861 | 2.4645 |
| 2.430 | 0.06677 | 2.4302 | 2.461 | 0.12623 | 2.4621 | 2.481 | 0.16209 | 2.4806 |
| 2.449 | 0.04289 | 2.4483 | 2.478 | 0.09790 | 2.4787 | 2.497 | 0.13304 | 2.4959 |
| 2.466 | 0.01604 | 2.4659 | 2.493 | 0.06853 | 2.4940 | 2.512 | 0.10263 | 2.5102 |
| 2.480 | -0.00734 | 2.4798 | 2.506 | 0.04320 | 2.5061 | 2.523 | 0.07553 | 2.5219 |
| 2.495 | -0.03641 | 2.4954 | 2.520 | 0.01231 | 2.5198 | 2.536 | 0.04221 | 2.5354 |
| 2.511 | -0.06774 | 2.5109 | 2.534 | -0.02131 | 2.5337 | 2.549 | 0.00740 | 2.5486 |
| 2.523 | -0.09483 | 2.5234 | 2.545 | -0.04962 | 2.5448 | 2.560 | -0.02316 | 2.5596 |
| | | | 2.558 | -0.08465 | 2.5579 | 2.572 | -0.05937 | 2.5721 |
| | | | | | | 2.583 | -0.09670 | 2.5844 |

The experimental data points, in Table 1, were fitted using the ORIGINPRO software and the best set of values of Eq. 5 was found with the reduced chi-square value of 0.783. This value ($\chi^2_{red.}$) indicates a very good fitting because the number of the freedom grade is 25 (8 parameters in the Eq. 5 and 33 IxV points) (Bevington, 1996). Table 2 shows the fitted parameter values and the standard deviations in parentheses.

Table 2 – Fitted parameters values of Eq. 5 with the standard deviations in parentheses.

| Parameter | Top Cell | Middle Cell | Bottom Cell |
|---|---|---|---|
| $I_L$(A) | 0.208(7) | 0.145(3) | 0.28(5) |
| $\Delta I_L$(A) | 0.0283(24) | 0.130(17) | 0.61(7) |
| $I_{01}.I_{02}.I_{03}$($A^3$) | 8.405(15)x$10^{-43}$ | | |
| $R_S$(Ω) | 0.186(8) | | |

The values of Table 2 are according with the expected values given in the A, B and C considerations.

## 6 - CONCLUSION

This paper describes a method to determine the photogenerated current of the individual junctions of the triple junction solar cell. The method requires a minimum of three different current-voltage curves, as is indicated in the work, to obtain a good set of fitted current-voltage parameters including the series resistance value. An individual fitting of each curve (each column of Table 1) was attempted, but the fitted parameters were not consistent.

For the first time, it is possible to directly extract the three photogenerated current values of the triple junction solar cell by direct analysis of current-voltage characteristic solely. The values of the bottom, middle and top junction photogenerated current are in close agreement with the expected values, which are estimated by the source light spectral irradiance.

The experimental apparatus to achieve the proposed task can be easily implemented at any solar cell characterization laboratory.

The method is largely applicable to all kinds of multi-junction solar cells and is very useful in studying the solar cells performance as a function of environmental parameters (temperature, radiation damage, thermal cycling, illumination system and others)

**ACKNOWLEDGMENTS**

The authors would like to acknowledge the FINEP (Financiadora de Estudos e Projetos), Brazilian governmental entity that promotes scientific and technological development, referent for the process CTEnerg III. The author (N.V.) is also grateful to OriginLab (www.originlab.com) by license to use his software in the data analysis of this work.